\documentclass[showpacs,preprint,amssymb]{revtex4}
\usepackage{graphicx}
\usepackage{dcolumn}
\usepackage{bm}
\def\be{\begin{equation}} 
\def\ee{\end{equation}}
\def\bea{\begin{eqnarray}} 
\def\eea{\end{eqnarray}}
\def\line{\hbox to \hsize}    
\def\frac #1#2{{#1\over #2}}

\def\tr{{\rm  tr\,}}

\def\ha{\hat a}
\def\had{\hat a^{\dagger}}
\def\hb{\hat b}
\def\hbd{\hat b^{\dagger}} 
\def\ad{a^{\dagger}}
\def\bd{b^{\dagger}}

\def\hH{{\hat H}}
\def\ad{ a^{\dagger}}

\def \ket #1{{\vert #1\rangle}}
\def \bra #1{{\langle #1\vert}}
\def \brak #1#2{{\langle#1\vert#2\rangle}}
\def\eval #1#2#3{{\langle#1\vert#2\vert#3\rangle}} 
\def\vev #1{{\langle #1\rangle}}
\def\1{\mbox{\bf 1}}

\begin{document}

\title{Mass flows  and angular momentum density for  $p_x+ip_y$ paired fermions in a harmonic trap}

\author{ MICHAEL STONE}

\affiliation{University of Illinois, Department of Physics\\ 1110 W. Green St.\\
Urbana, IL 61801 USA\\E-mail: m-stone5@uiuc.edu}   

\author{INAKI ANDUAGA}

\affiliation{University of Illinois, Department of Physics\\ 1110 W. Green St.\\
Urbana, IL 61801 USA\\E-mail: anduaga2@uiuc.edu}

\begin{abstract}  We present a simple   two-dimensional model of  a $p_x+ip_y$ superfluid in which the mass flow that gives rise to the intrinsic angular momentum  is  easily calculated by numerical diagonalization of the Bogoliubov-de Gennes operator. We find that, at zero temperature and for constant director $\bf l$, the  mass flow closely follows the Ishikawa-Mermin-Muzikar formula ${\bf j}_{\rm mass}=
\frac 12   {\rm curl\/} (\rho \hbar {\bf l}/2)$.

\end{abstract}

\pacs{ 71.10.Pm, 73.43.-f, 74.90.+n}

\maketitle

\section{Introduction}

Controversies over the  ``instrinsic angular momentum'' of the  A phase of superfluid  $^3$He have a long history \cite{kita1,leggett_book}, but the issue has contemporary relevance  for  the search for boundary currents in the suspected  $p_x+ip_y$ superconducting phase of Sr$_2$RuO$_4$ \cite{maeno_review}, and for  experiments on possible $p$-wave  versions of atomic Fermi condensates \cite{jin}.

Behind the competing computations  of the ground-state angular momentum density lie two competing intuitive pictures.  In the first the superfluid  is  regarded as a Bose condensate of Cooper pairs, each  pair  possessing   angular momentum  $\hbar$. In this  picture, and at temperature $T=0$, the total angular momentum of a spatially uniform system of ${\mathcal N}$ particles should be  $\frac 12 {\mathcal N}\hbar $. The second  picture begins with the observation that  in  the BCS regime, where $\Delta\ll \epsilon_F$, the  opening of  the energy gap $\Delta$  affects only  states  lying within an energy range of  a few $ \Delta$ about the  Fermi surface at $E=\epsilon_F$.   It is therefore anticipated  that the  ``na{\"\i}ve'' estimate of  $\frac 12 {\mathcal N}\hbar$ will  be  reduced by a factor of $\Delta/\epsilon_F$ \cite{anderson-morel}, or even by a factor of $(\Delta/\epsilon_F)^2$ \cite{cross1}.

For a system consisting of a fixed (even) number of particles with a common, angular momentum $l_z=1$   pair wavefunction $\psi({\bf r}_1,{\bf r}_2)$,  the  antisymmetric   many-body wavefunction is given by the pfaffian ${\rm Pf}[\psi({\bf r}_i,{\bf r}_j)]$, and  is undoubtedly an 
 eigenstate  of the angular momentum operator $L_z$ with eigenvalue $\frac 12 \hbar {\mathcal N}$ \cite{McClure}.  The problem is that $\psi({\bf r}_1,{\bf r}_2)$ is usually computed for a spatially uniform system, and  a  uniform  system is not suitable for computing angular momentum: a  substantial  contribution to the angular momentum can arise from a small current at large distances---in particular near the walls of the container,  where the density abruptly  drops to zero. 

Once we allow spatial variations of the fluid density or   order parameter, analytic expressions can only be obtained by  approximate methods. Unfortunately,  different approaches    have led to different answers (see  ref.\ \cite{kita1} for a brief summary). 
The Gorkov  gradient expansion, for example, computes   Green functions for   the Bogoliubov-de Gennes  (BdG) equation, and  in this  formalism the angular momentum arises  from the spectral asymmetry of the BdG differential operator. This asymmetry is closely related to the phenomenon of topologically induced fractional charge, and to the axial anomaly of QED.   Now the  axial anomaly is   quite a subtle phenomenon and can easily be missed if one makes na{\"\i}ve manipulations in conditionally convergent  integrals, or  appeals  to  symmetries that are vitiated by boundary conditions.  We claim  however, that when solved  exactly,  the  BdG equation  produces results entirely consistent  with the Cooper-pair wave function approach: there is no  $\Delta/\epsilon_F$ suppression, and the  ground-state  intrinsic angular momentum is $\frac12 \hbar$ per particle.  

In this paper we use the BdG equation to obtain  numerical results   for the angular momentum and associated mass-flows in  a two-dimensional model of $p_x+ip_y$ superfluid  fermions confined in a harmonic trap.  Numerical computations  of  the angular momentum in a cylindrical container have been obtained by Kita \cite{kita2}, and our results are entirely consistent with this  earlier work. We believe, however,  that the simplicity of the present model, where the phenomena  can be studied interactively with a few lines  of {\it Mathematica}$^{\rm TM}$  code,  make it of interest.  We begin in sections \ref{SEC:BdG} and \ref{SEC:2dHO} with a brief review of  the Bogoliubov de Gennes formalism and the two-dimensional Harmonic oscillator.  In section \ref{SEC:tridiagonal} we  show how symmetries and the harmonic oscillator selection rules serve to reduce the BdG equation  to  a tridiagonal matrix eigenvalue problem which is solved numerically in section \ref{SEC:numeric}. We end  in section \ref{SEC:discussion}  with a brief discussion of what we conjecture  to be the origin of the much smaller angular momentum found by some older methods.

 \section{Superfluidity  and the Bogoliubov-de Gennes equation}
 \label{SEC:BdG}

Because of the  history of controversy in this subject, it is essential that we explain exactly how we do our   calculations. We begin, therefore, with a brief review of the Bogoliubov-de Gennes formalism. 
 
Suppose that $H_{ij}$ is an  $N$-by-$N$ matrix representing a  one-particle ({\it i.e.\/}\ first quantized) Hamiltonian $H$. The second-quantized many-body hamiltonian corresponding to this is 
\be 
\hH = \had_iH_{ij}\ha_j,
\ee
where a sum over the repeated indices is to be understood.
Here  $\had_i$ and $\ha_i$ are second-quantization fermion creation and annihilation operators obeying
\be
\{\ha_i,\ha_j\}=\{\had_i,\had_j\}=0, \quad\{\ha_i,\had_j\}=\delta_{ij}.
\ee
The operator $\had_i$ creates a particle in state $i$ and $\ha_i$ destroys such a particle.
If  we represent $H$ in  a continuous basis, the index  ``$i$''  should be understood to incorporate both the space co-ordinate $x$, and any  spin index. A sum over $i$  therefore implies both an integral over real space and a sum over spin components.

 To account for   the effect of a condensate of   Cooper pairs, we allow  pairs of particles to disappear into or appear out of the background, 
 and  the second-quantized  $\hat H$ is replaced by the Bogoliubov Hamiltonian
\be
 \hat H_{\rm Bogoliubov}= \had_i H_{ij}\ha_j +{\textstyle\frac 12}  \Delta_{ij} \had_i\had_j +{\textstyle \frac 12} \Delta^{\dagger}_{ij} \ha_i \ha_j.
 \ee
 The  gap-function  matrix  $\Delta_{ij}$  is  skew symmetric. Different forms of $\Delta_{ij}$ describe different condensates and result in different patterns   of symmetry breaking.  The entries in   $\Delta_{ij}$ are  usually determined by a self consistency condition, or { gap equation\/}. In the present work we are more concerned with the consequences of a non-zero $\Delta_{ij}$ than with its origin, and so will take its magnitude and form as something imposed  externally.  
 
\begin{figure}
\includegraphics[width=5.0in]{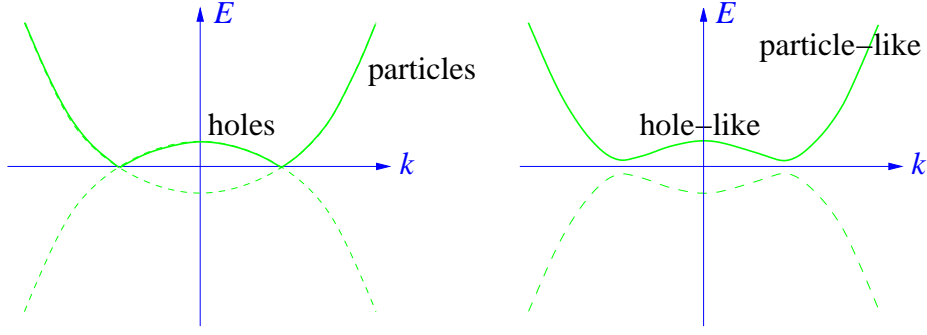}
\caption{The BdG operator  spectrum for a one-dimensional non-relativistic gas with \hbox{$E(k)=k^2/2-\mu$}. Left: the spectrum with $\Delta=0$.  Right: the spectrum after  a gap is opened by a non-zero $\Delta$.  In each case the part of the spectrum with BdG eigenvalue $E>0$ is shown as a solid line, and the part with $E<0$ is shown dashed. For $\Delta \ne 0$, the  $E>0$  ``particle-like'' region  has $|\vec v|\approx 0$ and $|\vec u|\approx 1$, and the ``hole-like'' region  has $|\vec u|\approx 0$ and $|\vec v|\approx 1$.  }
\label{FIG:BdG}
\end{figure}

  We can write the particle-number non-conserving Bogoliubov Hamiltonian as 
 \be 
 \hat H_{\rm Bogoliubov}= \frac12 \left(\matrix{ \had_i &\ha_i}\right)\left(\matrix{ H_{ij}& \phantom {-}\Delta_{ij}\cr
                                                                              \Delta^{\dagger}_{ij}& -H^T_{ij}}\right)
                  \left(\matrix{ \ha_j\cr \had_j}\right) +\frac 12 \tr H,                                                              
\ee
where  $H^T$ denotes the transpose of the hermitian matrix $H$.  
The two-by-two block-matrix form of the   many-body  Hamiltonian conveniently  allows it to be diagonalized  by means of a  Bogoliubov transformation.  We  first solve  the single-particle  Bogoliubov-de-Gennes (BdG) eigenvalue problem
\be
 \left(\matrix{ H & \Delta\cr
                        \Delta^{\dagger}& -H^T}\right)
                  \left(\matrix{ \vec u_m \cr  \vec v_m}\right)=E_m  \left(\matrix{ \vec u_m \cr  \vec  v_m}\right).
 \label{EQ:eigenvalue}
 \ee                 
 Here  $\vec u_m$ and  $\vec v_m$ are $N$-dimensional column vectors, which we take to be normalized so that $|\vec u_m|^2+|\vec v_m|^2=1$. If we explicitly  display the  column-vector index $i$, these vectors  can be regarded as $N$-by-$N$ matrices with entries $u_{im}$
 and $v_{im}$.  
Taking the complex conjugate  of (\ref{EQ:eigenvalue}) tells us  that
\be
 \left(\matrix{ H & \Delta\cr
                        \Delta^{\dagger}& -H^T}\right)
                  \left(\matrix{\vec  v^*_m \cr  \vec u^*_m}\right)=-E_m  \left(\matrix{ \vec v^*_m \cr  \vec u^*_m}\right),
\label{EQ:negative_eigenvalue}                 
 \ee       
 and so the BdG-operator  eigenvalues come in $\pm$ pairs. We will   always use the symbol   $E_m$ to refer to  the positive eigenvalue, and a sum over $N$ otherwise unspecified  eigenvectors  is a sum over the positive $E_m$ eigenvectors. The manner in which the BdG operator doubles  the spectrum, and how we  retain  only the positive part, is illustrated in Figure \ref{FIG:BdG}.
 
 We now  set  
 \bea
 \ha_i&=& u_{im}\hb_m +v^*_{im}\hbd_m\nonumber\\
 \had_i&=& v_{im} \hb_m +u^*_{im}\hbd_m.
 \eea
 The mutual orthonormality  and completeness of the eigenvectors $(\vec u_m, \vec v_m)^T$ ensures that the $\hb_m$, $\hbd_m$ have the same anti-commutation relations as the $\ha_i$, $\had_i$. In terms of the $\hb_m$, $\hbd_m$, the second-quantized Hamiltonian becomes   
 \be
 \hat H_{\rm Bogoliubov} =\sum_{m=1}^N E_m \hbd_m \hb_m -\frac 12 \sum_{m=1}^N E_m +\frac 12\sum_{i=1}^N E^{(0)}_i.
\label{EQ:bogdiag}
\ee
Here the $E^{(0)}_i$ are the eigenvalues of $H_{ij}$. Unlike the $E_m$, these can be of either sign.
If all the quasi-particle excitation energies $E_m$ are strictly positive, the  new ground state is non degenerate and is the unique state $\ket{0}_b$   annihilated by all the $\hb_m$.   The ground-state expectation value  of an  operator $\hat O=\had_iO_{ij}\ha_j$ is therefore 
\bea
\vev{\hat O}&=&{}_b\bra{0}\had_iO_{ij}\ha_j\ket{0}_b\nonumber\\
 &=& _b\eval{0}{(v_{im} \hb_m +u^*_{im}\hbd_m)O_{ij}(u_{jn}\hb_n +v^*_{jn}\hbd_n)}{0}_b\nonumber\\
&=&  v_{im}O_{ij}v^*_{jm}\nonumber\\
&=& \sum_{m=1}^N \vec v_m^T O\vec v^*_m.
\eea 
Because of the $E\leftrightarrow -E$, $\vec u\leftrightarrow \vec v^*$ symmetry, we  could equivalently write this  last expression  as   $\sum \vec u_m^\dagger O\vec u_m$, the sum being  taken over the  $N$ {negative-energy\/} eigenstates of $\hH_{\rm Bogoliubov}$, which we think of as being a   filled  Dirac sea.

For example, the average  number of particles present in the system is found by taking $O_{ij}=\delta_{ij}$,
and is
\be
{\mathcal N}= {_b}\bra{0}{\had_i\ha_i}\ket{0}_b= v_{im}v^*_{im}= \sum_{m=1}^N  |\vec v_m|^2=\sum_{{\rm negative}\,E} |\vec u_m|^2.
\ee
  When $\Delta\equiv 0$,  this sum will be  equal to the number of negative-energy eigenstates of $H_{ij}$.

Although we will not make  use of it, it is worth pointing out that if we define the  matrix 
\be
\psi_{ij}=   v^*_{im}(u^{-1})^*_{m j},
\ee
then then the orthogonality and completeness conditions tell us that $\psi_{ij}=-\psi_{ji}$. 
The  new ground-state $\ket{0}_b$ can be written in terms of $\psi_{ij}$ as
\be
\ket{0}_b =C  \exp\left\{{\textstyle \frac 12} \had_i\had_j\psi_{ij}\right\} \ket{0}_a.
\label{EQ:ketb}
\ee
 Here $\ket{0}_a$ is the vacuum state  annihilated by all the $\ha_i$, and $C$ is a normalization constant.
This result  exhibits the ground state  as a coherent superposition of paired states, with $\psi_{ij}$ being the un-normalized Cooper-pair wavefunction. The state $\ket{0}_b$  gives  a non-zero expectation value for the fermion-number non-conserving order parameter 
\be
_b\bra{0}{\ha_i\ha_j}\ket{0}_b=\sum_n u_{in} v^*_{jn},
\ee
and so possesses a sharp order-parameter phase and an uncertain  particle number. If we desire to work with a fixed (even) number of particles ${\mathcal N}$, we should retain only the $\frac 12 {\mathcal N}$-th term in the series expansion of the exponential in equation (\ref{EQ:ketb}).  The resulting many-body wavefunction is then the pfaffian
\be
\Psi(i_1,\ldots,i_{{\mathcal N}})\stackrel{\rm def}{=}{}_a\bra{0}{\ha_{i_1}\cdots \ha_{i_{\mathcal N}}} \ket{{\mathcal N}}= {\rm Pf}[\psi_{i_\alpha,i_\beta}].
\ee
In the large $\mathcal N$ limit  there should be no locally measurable physical
distinction between the sharp-phase and the sharp-particle-number ground states.

\section{Two-dimensional Harmonic oscillator}
\label{SEC:2dHO}

We consider a  model that  consists of   spinless fermions confined in a harmonic trap in the $x$-$y$ plane.  The one-particle Hamiltonian $H$ of section \ref{SEC:BdG} is therefore that of the two-dimensional harmonic oscillator:
\be
H= \frac 12 \left(p_x^2+p_y^2+ \omega^2(x^2+y^2)\right). 
\ee
Here, for example, $p_x=-i\hbar \partial/\partial x$ and so  $ [x,p_x]=i\hbar$. We have taken the  mass of the trapped   particles to be unity.  We define  the usual harmonic-oscillator ladder  operators 
\be
a_x = \sqrt{\frac{\omega}{2\hbar}}\left(x+i\frac {p_x}{\omega}\right),\quad a_x^\dagger= \sqrt{\frac{\omega}{2\hbar}}\left(x-i\frac {p_x}{\omega}\right),
\ee
and similarly $a_y$ and $a^\dagger_y$.
These operators obey 
\be
[a_x,\ad_x]= [a_y,\ad_y]=1,\quad [a_x,a_y]=[\ad_x,a_y]=0,
\ee
and, in terms of them,
\be
H= \hbar \omega (a_x^\dagger a_x+a_y^\dagger a_y +1).
\ee
The normalized  eigenstates can be written as  
\be
\ket{n_x,n_y}= \frac{(\ad_x)^{n_x}}{\sqrt{n_x!}}  \frac{(\ad_y)^{n_y}}{\sqrt{n_y!}}\ket{0,0}, 
\ee
and have energy eigenvalues $E_{n_x,n_y}=\hbar\omega(n_x+n_y+1)$. 

\begin{figure}
\includegraphics[width=4.0in]{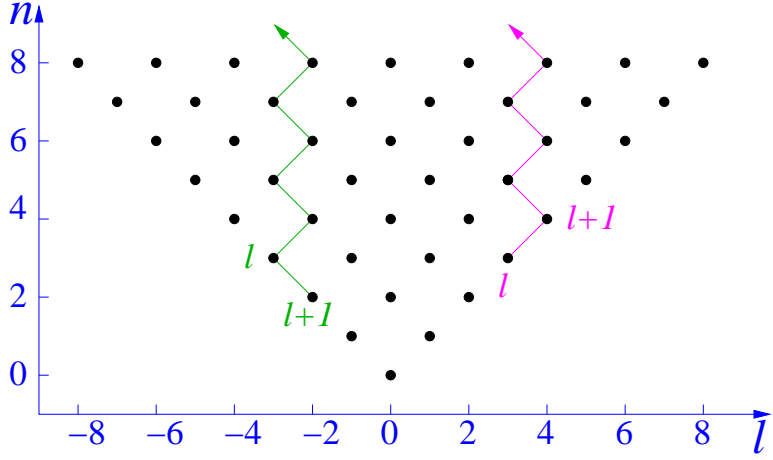}
\caption{The low-lying part of the  harmonic oscillator spectrum. The zig-zag paths indicate how  states are coupled by the tridiagonal matrix $H^{(l)}$.  For $l$ positive, the first entry in the eigenvector will be a ``$v$'', and the second entry a ``$u$,'' and so on. For $l$ negative, the $u$'s and $v$'s are interchanged.}
\label{FIG:trid}
\end{figure}

We will find it more useful to work in a basis in which the angular momentum operator
\be
L_z\equiv xp_y-yp_x=ih(\ad_y a_x -\ad_x  a_y)
\ee
is diagonal with eigenvalues $\hbar l$. To construct  this basis, we define new  ladder  operators
\be
b_1^\dagger =\frac{1}{\sqrt{2}}(a_x^\dagger+ia_y^\dagger),\quad b_2^\dagger =\frac{1}{\sqrt{2}}(a_x^\dagger-ia_y^\dagger),
\ee
which obey
\be
[L_z,b_1^\dagger]=\hbar b_1^\dagger, \quad[L_z,b_2^\dagger]= -\hbar b_2^\dagger.
\ee
Consequently  $\bd_1$ increases the angular momentum quantum number $l$  by unity and $\bd_2$ decreases it by unity. It terms of these new operators  we have 
\be
H= \hbar \omega (b^\dagger_1b_1+b_2^\dagger b_2 +1),
\ee
and the eigenstates become 
\be
\ket{n,l}= \frac{(b_1^\dagger)^N}{\sqrt{N!}} \frac{(b_2^\dagger)^M}{\sqrt{M!}}\ket{0,0},
\quad n=N+M, \quad l=N-M.
\ee
The angular momentum of the state $\ket{n,l}$ is $\hbar l$ and its energy is 
$
 E_{n,l}= \hbar\omega (n+1).
$
The set of states with energy quantum number $n$ is $(n+1)$-fold degenerate, the angular-momentum quantum number $l$ running from $l=-n$ to $l=n$ in steps of two.  (See Figure \ref{FIG:trid}.)

We will require the normalized  real-space wavefunctions $\brak{r,\theta}{n,l}$ corresponding to these eigenstates.
These wavefunctions are commonly  written as 
\be
 \psi_{N,l}(r,\theta) \equiv  \brak{r,\theta}{2N+|l|,l} =\omega^{|l|+1/2} \sqrt{\frac{N!}{\pi(N+|l|)!}} e^{il\theta} r^l e^{-\omega r^2/2}L^{|l|}_N(\omega r^2),
 \ee
 where $L^{|m|}_n(x)$ is the associated Legendre polynomial 
 \be
 L^{|m|}_n(x) = \frac{x^{-|m|} e^x}{n!} \frac{d^n}{dx^n} e^{-x} x^{n+|m|},
\ee
which is  of degree $n$. As the notation  $\brak{r,\theta}{2N+|l|,l}$ indicates, these  states have energy 
\be
 E_{N,l}=\hbar\omega(2N+|l|+1),
 \ee
 and so the integer  $N$ counts the height   (starting from zero at the lowest point) of the state in  the column of eigenstates of angular momentum $l$.
 
 \section{Bogoliubov-de Gennes eigenstates}
 \label{SEC:tridiagonal}

We now suppose that some suitable interaction has caused our  fermions to enter a superfluid phase characterized by an order parameter with  $p_x+ip_y$ symmetry.   We can think of the fluid as being a single atomic layer of $^3$He in the A-phase and with the angular momentum director $\bf l$ of the Cooper pairs pointing in the $+\hat z$ direction. 
  We  therefore  wish to diagonalize  the  Bogoliubov-de Gennes  operator 
 \be
 H_{\rm BdG}=\left(\matrix{H-\mu & \Delta(p_x+ip_y) \cr \Delta (p_x-ip_y)& -(H-\mu)}\right).
 \ee
Here $H$ is the harmonic oscillator hamiltonian, $\mu$ is a chemical potential that  controls the number of particles in the trap, and $\Delta$ is a scalar parameter. The off-diagonal term
 \be
\Delta( p_x+ip_y) = i\Delta \sqrt{\hbar\omega} (b_1^\dagger -b_2)
 \ee  
 increases the angular momentum quantum number of any state on which it acts  by unity, and 
 represents the effect of  the  condensate of Cooper pairs, each pair  possessing angular momentum $+\hbar \hat z $. 
 
 The  eigenstates with eigenvalue  $E=E_{m,l}$    will be of the form 
 \be
 \Psi_{m,l}(r,\theta) = \left[\matrix{iu_{m,l}(r,\theta)\cr v_{m,l}(r,\theta)}\right]= \left[\matrix{i\sum_n u_{m,l}^{n} \brak{r,\theta} {n,l+1}\cr \sum_n v_{m,l}^{n} \brak{r,\theta}{ n,l}}\right].
 \ee
 The sum over $n$ is over all harmonic-oscillator states consistent with the $l$ quantum number. The pair $(n,l)$ corresponds to the index ``$i$'' in the general theory in section \ref{SEC:BdG}. 
 We choose to label the BdG eigenstates by the angular momentum  of the lower component ``$v$.''  Because of the unit off-set between the angular momentum of the $u$'s and $v$'s, the $E\leftrightarrow -E$ pairing is between eigenstates states with angular-momentum label $l$ and those with label $-l-1$. 
 
In the harmonic-oscillator eigenstate basis, and for any fixed $l$, the BdG operator reduces to a tridiagonal matrix connecting  only the 
$l$ and $l+1$   subspaces.  The pattern of $H_{\rm BdG}$-connected states is shown in Figure 
\ref{FIG:trid} for both positive and negative $l$. This  pattern depends on the sign of $l$ because, for  $l$ is negative, the ladder operator $b_2$ appearing in $p_x+ip_y$ acts non-trivially on $\ket{|l|,-|l|}$ to yield a state with lower energy. When $l$ is positive, however, it annihilates  $\ket{l,l}$. The factor of $i$ in the upper entry  of the column vector $\Psi_{m,l}$ has been inserted  to eliminate   $i$'s and  minus signs in the tridiagonal matrix, so that it becomes a real symmetric matrix with positive off-diagonal terms.

The necessary  matrix elements can be read-off from   
\bea
i(\bd_1-b_2)\ket{n,l}&=& i\sqrt{\frac{n+l+2}{2}}\ket{n+1,l+1}-i\sqrt{\frac{n-l}{2}}\ket{n-1,l+1}\nonumber\\
i(\bd_2-b_1)\ket{n,l+1}&=& i\sqrt{\frac{n-l-1}{2}}\ket{n+1,l} -i\sqrt{\frac{n+l+1}{2}}\ket{n-1,l}\nonumber\\
(H-\mu)\ket{n,l}& =&\epsilon_n\ket{n,l},
\eea
where
$$
\epsilon_n =\hbar\omega(n+1)-\mu. 
$$
For 
$l$ positive or zero, and with $h\omega =1$, the $(l,l+1)$-subspace BdG matrix takes the form,
\be
H^{(l)}=\left[\matrix{-\epsilon_l &\Delta\sqrt{l+1}& & & & &\cr
                    \Delta \sqrt{l+1}& \epsilon_{l+1} &\Delta\sqrt{1} & & & &\cr
                         &\Delta \sqrt{1}& -\epsilon_{l+2} &\Delta\sqrt{ l+2} & &&\cr
                         & &\Delta \sqrt{l+2} &\epsilon_{l+3}&\Delta \sqrt{2} &\cr
                         & & &\Delta  \sqrt{2}& -\epsilon_{l+4} & \ddots \cr
                        & & & & \ddots&\ddots \cr } \right].
 \ee       
 All entries more than one step away from the diagonal are zero.
When $l$ is strictly negative the $(-|l|,-|l|+1)$-subspace  BdG matrix  becomes 
\be
H^{(l)}= \left[\matrix{\epsilon_{-l-1} &\Delta\sqrt{-l}& & & & &\cr
                     \Delta\sqrt{-l}& -\epsilon_{-l} &\Delta\sqrt{1} & & & &\cr
                         & \Delta\sqrt{1}& \epsilon_{-l+1} &\Delta \sqrt{ -l+1} & &&\cr
                         & & \Delta\sqrt{-l+1} &-\epsilon_{-l+2}&\Delta \sqrt{2} &\cr
                         & & & \Delta \sqrt{2}& \epsilon_{-l+3} & \ddots \cr
                        & & & & \ddots&\ddots \cr } \right].
 \ee 
In either  case,  the $i,j$-th  matrix element  links states  with $i$,$j$  labeling the distance from the lowest point along the zig-zag paths shown in Figure 1.  The entries in the $m$-th eigenvector of $H^{(l)}$  will therefore consist of alternating $u_{m,l}^{n}$'s and $v_{m,l}^{n}$'s, the first entry being a ``$v$'' when $l$ is positive, and a ``$u$'' when $l$ is negative.

It is an easy task to numerically diagonalize a tridiagonal matrix, and so the effect of the gap parameter
$\Delta$ on the spectrum can be explored.

\section{Numerical Results}
\label{SEC:numeric}

We have computed \cite{stone-nb}  the spectrum  and the eigenvectors  $u_{m,l}^n$ and $v_{m,l}^n$ for a variety of values of $\mu$ and  $\Delta$. Typical results are displayed in this section. In all these plots we have set $\hbar\omega$ to unity. Changing the value of $\omega$ serves only to  rescale the energy and $r$. 

\begin{figure}
\includegraphics[width=5.5in]{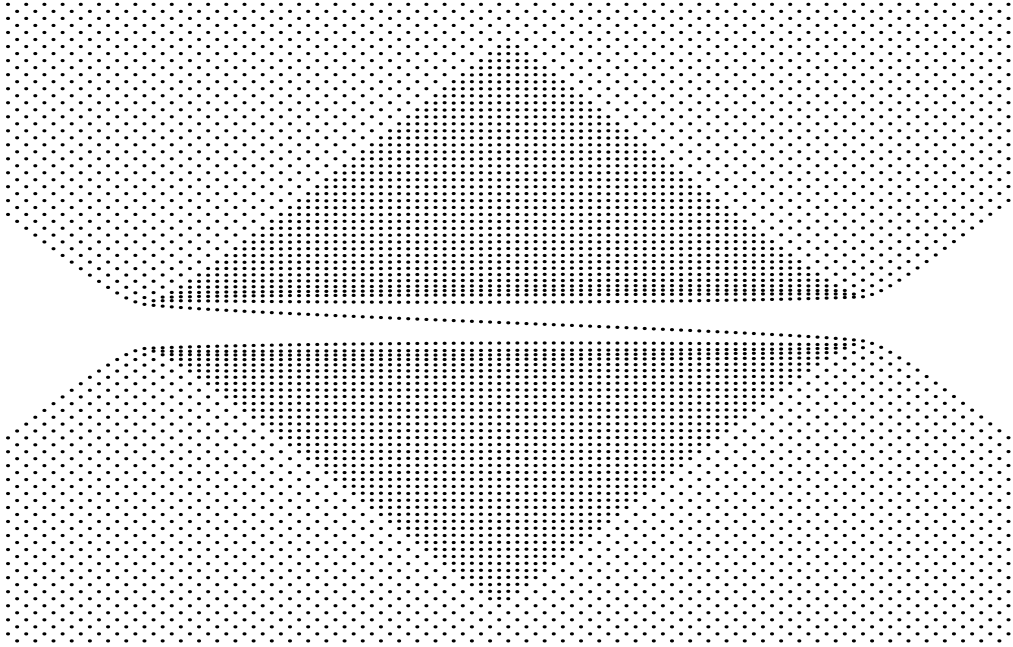}
\caption{Part of the BdG operator eigenvalue spectrum  for $\mu=40.1$  and $\Delta=0.5$. We have suppressed the axes for clarity. The horizontal co-ordinate,  the angular momentum eigenvalue $l$, runs from $-55$ to $+55$, and the eigenvalues $E_{m,l}$ are plotted vertically.  Each column of eigenvalues is  the result of diagonalizing a 100-by-100 tridiagonal matrix.}
\label{FIG:BdGspectrum}
\end{figure}

Figure \ref{FIG:BdGspectrum} shows a plot of the eigenvalues for the case   $\mu=40.1$ and $\Delta=0.5$. The interpenetrating wedges of  $\pm E$ copies of the harmonic oscillator spectrum are clearly visible, as is the gap lying symmetrically about about  $\epsilon=0$. The  family of states crossing the gap from the upper ``continuum''   to the lower  as $l$ increases is  a chiral Majorana  edge mode   whose existence was first pointed out by Volovik \cite{volovik-edge}, and whose importance for the analogy with the Pfaffian quantum Hall state was stressed by Read and Green \cite{read-green}. 

\begin{figure}
\includegraphics[width=4.5in]{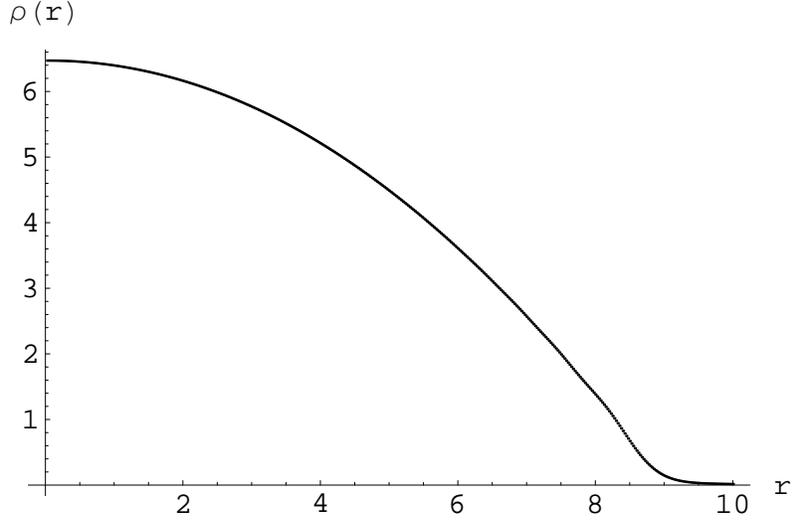}
\caption{Fluid density $\rho(r)$ as a function of radius for the case $\mu=40.1$, $\Delta=0.5$. The number of particles in the fluid is  ${\mathcal N}= 2\pi \int_0^\infty  \rho(r)\,rdr=840$.}
\label{FIG:density}
\end{figure}

We also show a plot (Figure \ref{FIG:density})  of the fluid density 
\be
\rho(r) = \sum_{l,m} |v_{m,l}(r,\theta)|^2
\ee
as a function of radius $r$. The $\Delta=0.5$ density profile differs  from that for $\Delta=0$ only by being slightly smoother. It closely follows the Thomas-Fermi density  
\be
\rho_{\rm TF}(r)= \cases{ \frac 1{2\pi} \left(\mu-\frac 12 r^2\right), & $r< \sqrt{2\mu},$\cr
                                           0, & $r> \sqrt{2\mu}.$}
\ee
The Thomas-Fermi approximation  estimates the  particle number to be 
\be
{\mathcal N}_{\rm TF}=2\pi \int_0^{\sqrt{2\mu}} \rho_{\rm TF}(r)\,rdr= \frac 12 \mu^2.
\ee
 The actual particle number for the free-particle case $\Delta=0$ is given by
${\mathcal N}_0 = \frac 12 \lfloor \mu\rfloor (\lfloor \mu \rfloor +1)$, where $\lfloor \mu \rfloor$ indicates the integer part of $\mu$. When $\Delta$ becomes non-zero   the particle-hole asymmetry in the harmonic oscillator density of states causes the particle number at fixed $\mu$ to creep upwards. The effect is small  for small $\Delta$, however.  

\begin{figure}
\includegraphics[width=4.5in]{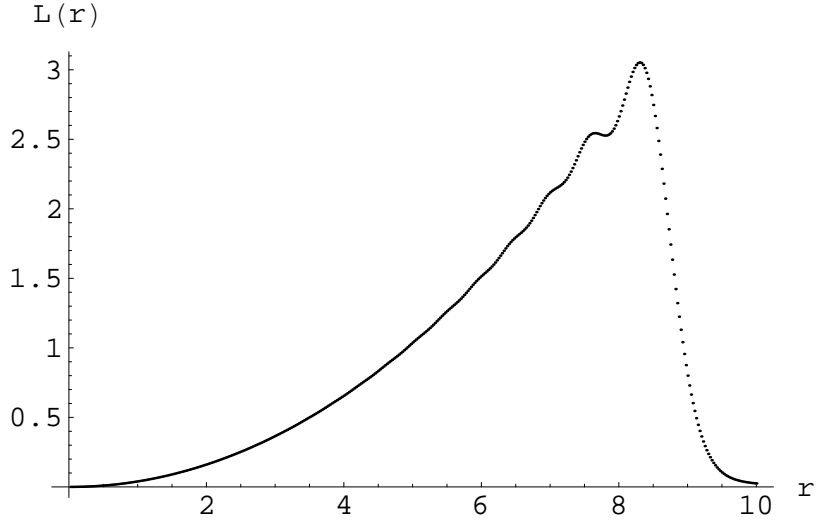}
\caption{Angular momentum density $L(r)$ for the same parameters as Figure \ref{FIG:density}. The total angular momentum is $L_{\rm tot}= 2\pi \int_0^\infty  L(r)\,rdr=410$. Thus, for these parameters, $L_{\rm tot}/{\mathcal N}=0.49$.}
\label{FIG:angmomdensity}
\end{figure}

\begin{figure}
\includegraphics[width=4.5in]
{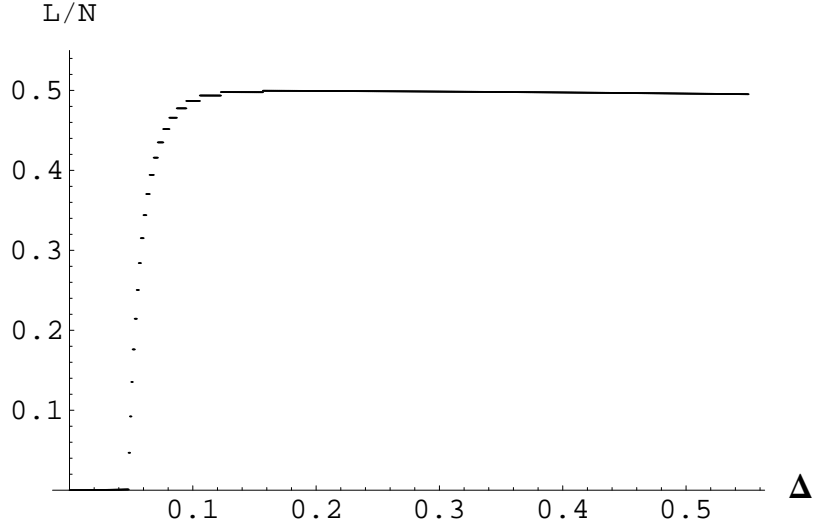}
\caption{The ground state angular momentum per particle $L_{\rm tot}/{\mathcal N}$  plotted as a function of $\Delta$ for $\mu=40.1$. The discontinuous  steps in the rising part of the graph are due to individual levels crossing $\epsilon=0$, and so changing their occupation. The  $\Delta>0.05$ ground state is therefore not adiabatically connected to the $\Delta=0$ ground state.}
\label{FIG:LNvsGap}
\end{figure}

\begin{figure}
\includegraphics[width=4.5in]{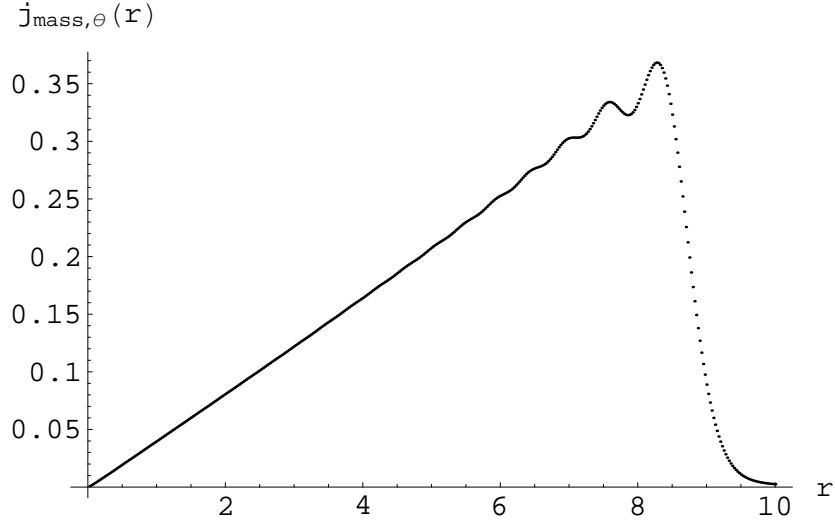}
\caption{The ground state azimuthal mass flow $j_{{\rm mass},\theta}(r)=L(r)/r$ corresponding to the angular momentum density in Figure \ref{FIG:angmomdensity}.}
\label{FIG:massflow}
\end{figure}

Figure \ref{FIG:angmomdensity} plots the ground state angular momentum density 
\be
L(r) =\vev{\hat L_z}= \sum_{l,m}  v_{m,l}\left(-i  \frac{\partial}{\partial \theta}\right) v^*_{m,l}=-\sum_{l,m} l |v_{m,l}(r,\theta)|^2  
\ee
  as a function of $r$ for the same parameters, and Figure  \ref{FIG:LNvsGap} shows the angular momentum per particle as a function of $\Delta$.
 Except for very small $\Delta$ we  find that
 \be
 L_{\rm tot} =2\pi \int_0^\infty L(r)\,rdr \approx  \frac 12 {\mathcal N}. 
\ee
There is no simple identity lying behind this fact, and mathematically it  results from a quite non trivial rearrangement of spectral weight between the positive and negative $E$ eigenstates of any given $l$. 
Physically, however, the   angular momentum arises  from an azimuthal  mass flow 
\be
j_{{\rm mass},\theta}(r)= \frac 1{2i} \sum_{l,m} \left\{ v_{m,l}\left(\frac 1 r \frac{\partial}{\partial \theta} v^*_{m,l}\right)-\left( \frac 1r \frac{\partial}{\partial \theta} v_{m,l}\right) v^*_{m,l}\right\}= \frac 1 r L(r).
\ee
This quantity is plotted in Figure \ref{FIG:massflow}. The  straight line mass-flow distribution is not fundamental, but indicates that the mass current is well described by the Ishikawa-Mermin-Muzikar formula \cite{mermin-muzikar,ishikawa0,ishikawa}  
\be 
{\bf j}_{\rm mass}=  \frac 1 2 {\rm curl\,} \left(\frac 12 \rho \hbar {\bf l}\right),
\label{EQ:ishikawa}
\ee
and that the fluid has a  near-parabolic particle density profile. 
 The Friedel-like oscillations near the abrupt drop of $j_{{\rm mass},\theta}(r)$ to zero at the edge of the droplet of confined fluid are also seen in the numerical results of ref.\ \cite{kita2}.

If the mass current  ${\bf j}_{\rm mass}$ is indeed given by   (\ref{EQ:ishikawa}), then an integration by parts   gives 
\be
L_{\rm tot} = \int ({\bf r}\times {\bf j}_{\rm mass})\,  d^3 {\bf r}= \frac 1 2 \hbar \int \rho\,{\bf l}\, d^3{\bf r}= \frac 12 {\mathcal N}\hbar {\bf l}.
\ee
The quantity $\frac 12 \rho \hbar {\bf l}$ can therefore  be identified with an intrinsic angular momentum density. The current associated with  this angular momentum density is then analogous to the bound electric current 
$
{\bf J}_{\rm bound} = {\rm curl\,} {\bf M}
$
associated with a  magnetic-moment density  $\bf M$.  

It is interesting to ask how much of the mass flow and angular momentum is  supplied  by the chiral Majorana edge mode. This branch of states is, after all, the most strikingly $l \leftrightarrow -l$ asymmetric feature of the BdG spectrum.  A first (but misleading) estimate suggests that  these states  account for   the entirety of the angular momentum.   Figure \ref{FIG:BdGspectrum} shows that   positive-energy  within-gap states  exist for each integer  $l$ in the range  $-\mu$ to $0$.  Because they are confined near the fluid boundary  by Andreev reflection, each of these  states  consists of an equally-weighted  linear superposition of   particle and hole,  and so has $|\vec u|^2=|\vec v|^2= {\textstyle \frac12}$. They therefore   contribute an angular momentum of  
$$
L_{\rm tot}^{\rm edge-mode}= -\frac 12 \sum^{0}_{l=-\mu} \hbar l = \frac 1  4 {\mu}(\mu+1)\hbar \approx \frac 12 {\mathcal N}\hbar = L_{\rm tot}.
$$
This result should be contrasted with  what happens in Kita's model of  a uniform fluid bounded by a rigid wall  \cite{kita2}. In this  case  the edge modes have dispersion $E(k)= -  \Delta (k/k_F)$ \cite{stone-roy} and merge into the continuum at $k=k_F$.  They therefore contribute   a boundary  current---equivalently a   momentum per unit length---of magnitude 
$$
j_{\rm boundary}^{\rm edge-mode} = -\frac 12 \int^0_{-k_F}\hbar k_z \frac{dk_z}{2\pi}  =  \frac \hbar {8\pi} k_F^2 = \frac{\hbar }{2} \rho_{\rm bulk},
$$
where $\rho_{\rm bulk}= k_F^2/4{\pi}$ is the bulk fluid density.  This current is larger  by a factor of two  than the boundary  current $j_{\rm boundary}=\hbar \rho_{\rm bulk}/4$  obtained from ({\ref{EQ:ishikawa}) by making $\rho$ decrease  from  $\rho_{\rm bulk}$ to zero   
as we approach  the container wall.  The rigid-wall edge modes therefore oversupply  angular momentum by a factor of two.  It  was shown in \cite{stone-roy} that this  twice-too-large bound state angular momentum  is reduced  by contributions from the unbound continuum states, and that the resulting  edge momentum density is exactly what is required to give the ${\mathcal N}\hbar/2$ total angular momentum.

\begin{figure}
\includegraphics[width=4.5in]{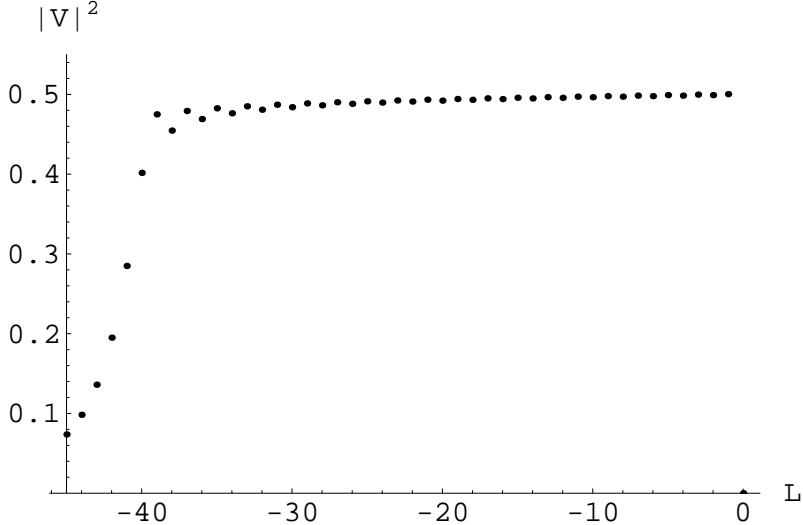}
\caption{The coeffecient $|\vec v|^2$  for the lowest positive energy modes. The point at which the edge-modes merge into the upper continuum is signalled by the sharp decrease in $|\vec v|^2$ near $l=-40$. The parameters are the same as those in  Figure   \ref{FIG:density}. }
\label{FIG:boundv2}
\end{figure}

\begin{figure}
\includegraphics[width=4.5in]{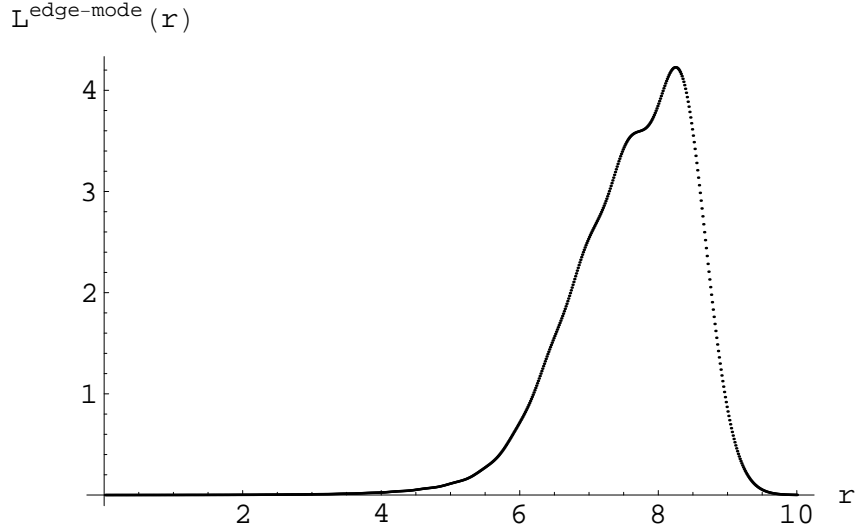}
\caption{The contribution of   edge modes with  $-41\le l<0$ to the angular momentum density $L(r)$.  The  parameters are those of  Figure \ref{FIG:density}.   }
\label{FIG:boundangmomdensity}
\end{figure}

\begin{figure}
\includegraphics[width=4.5in]{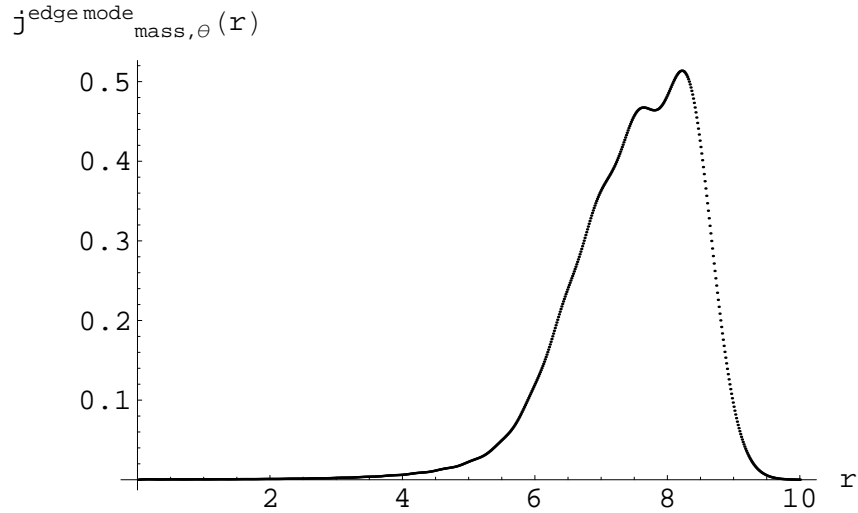}
\caption{The $-41\le l<0$ edge mode contribution to the mass flow. The   parameters are the same as those in  Figure \ref{FIG:density}. }
\label{FIG:boundmassflow}
\end{figure}

To investigate whether the edge modes are the source of  the {\it entire\/}  harmonic trap edge current, we have isolated   their contribution to the angular momentum density, and to the mass-flow current. 
Figure \ref{FIG:boundv2} shows that it is easy to determine which of the low-lying positive energy states states should be considered  Andreev bound-state edge modes. When we sum  the contributions to the angular momentum density and the mass flow from  these states only, we obtain the results shown in Figures
 \ref{FIG:boundangmomdensity} and \ref{FIG:boundmassflow}.  Although 
 \be
 2\pi \int_0^\infty L^{\rm edge-mode}(r)\,rdr\approx 2\pi \int_0^\infty L(r)\,rdr 
 \ee
 as anticipated,  both $L^{\rm edge-mode}(r)$ and $j_{{\rm mass},\theta}^{\rm edge-mode}(r)$ are   
  localized near the boundary of the fluid, and differ substantially from $L(r)$ and $j_{{\rm mass},\theta} (r)$.  We conclude  that,  as with the rigid wall model of \cite{kita2,stone-roy}, the continuum modes provide  an important component   of the mass-flow.
 That the  bound-state angular momentum contribution turns out to be  equal to the total angular momentum should therefore be regarded as a coincidence arising from the particular form of the harmonic trap density profile.

\section{Discussion}
\label{SEC:discussion}

There are a number of gradient-expansion results for the mass current in three-dimensional $^3$He-A.
By   mapping the problem onto one involving  fractional charge \cite{goldstone-wilczek},  and for a uniform  mass  mass-density $\rho_0$, Garg {\it et al.\/}\  \cite{stone-garg-muzikar}  obtained
\be
{\bf j}_{\rm mass}=\rho_0 {\bf v}_s + \frac1 4  \rho_0\,{\rm curl\,} \hbar {\bf l} -\frac 1 2 \rho_0\,{\bf l}
({\bf l}\cdot {\rm curl\,} \hbar {\bf l})
\label{EQ:cross-garg}
\ee
where ${\bf v}_s$ is the   superfluid velocity. This expression coincides with that obtained under the same conditions in \cite{cross1}.  Because it does not allow for  variations in the density, it cannot be used to compute the boundary currents.
Mermin and Muzikar \cite{mermin-muzikar} used a more sophisticated gradient expansion and  found that  when $\rho$ is allowed to vary slowly on the length scale of the coherence length we have
\be 
{\bf j}_{\rm mass}= \rho {\bf v}_s+\frac 1 4 {\rm curl\,} \rho \hbar {\bf l}- \frac 1 2 c_0\,{\bf l}
({\bf l}\cdot {\rm curl\,} \hbar {\bf l}).
\label{EQ:mermin-muzikar}
\ee
Here $c_0$ is a number that in the BCS  limit $\Delta\ll \epsilon_F$ is close to $\rho$, but goes to zero in the limit of tightly-bound Cooper pairs.  The  last term in both (\ref{EQ:cross-garg}) and
 (\ref{EQ:mermin-muzikar})  is known as the the {\it twist term\/}. The current associated with it is  now understood to be history dependent: if we start in the ground state  with $\bf l$ spatially constant  and then adiabatically deform  $\bf l$   to the desired texture, the twist-term  contribution to the mass flow  is cancelled by momentum  carried by excitations that have been forced through the nodes of the gap  by spectral flow (a manifestation of the QED axial anomaly) \cite{volovik-flow,balatskii-volovik-konyshev, stone-gaitan,volovik-texture}.

For a spatially constant director field $\bf l$,  and when ${\bf v}_s=0$, the Mermin-Muzikar result reduces to the  earlier formula obtained by Ishikawa {\it et al.\/} \cite{ishikawa0,ishikawa}: 
\be 
{\bf j}_{\rm mass}=  \frac 1 2 {\rm curl\,} \left(\frac 12 \rho \hbar {\bf l}\right).
\label{EQ:ishikawa1}
\ee
As we  described in section \ref{SEC:numeric}, although it is derived only for slowly varying $\rho$, our numerical results fit the two-dimensional version of (\ref{EQ:ishikawa1}) rather well. In particular, the striking straight-line dependence of ${j}_{{\rm mass},\theta}(r)$  is an immediate  consequence of   the inverted-parabola Thomas-Fermi  density profile being a good fit to the actual density.  

Since many computations of  the boundary   current and resulting angular momentum are in agreement in their common domain of applicability, why do other   estimates of the angular momentum,  such as \cite{anderson-morel,cross1},  find   results that are suppressed by powers of $\Delta/\epsilon_F$?   One explanation   is that one can cast \cite{stone-gaitan,stone-roy} the problem of the computing the boundary  current into   a weighted sum over $k_z$ of the quantity 
\be
j(k_z)= \lim_{s \to 0}\left\{- \frac 12 \sum_n {\rm sgn\,} (E_{n,k_z})|E_{n,k_z}|^{-s}\right\},
\label{EQ:oldedge}
\ee
where $E_{n,k_z}$ are the energy eigenvalues of a   Dirac hamiltonian     
 \be
H_{\rm Dirac}= -i\sigma_3\partial_x+\sigma_2k_z +m(x)\sigma_1
\ee
in which  $m(x)$ changes sign as $x$ passes through zero.
Now the operator 
\be
{Q}=\sigma_2 H_{\rm Dirac}-k_z
\ee
obeys 
\be 
{ Q}H_{\rm Dirac}=-H_{\rm Dirac}{Q}, 
\ee
and so it {\it seems\/}   that  all  the  eigenstates of $H_{\rm Dirac}$,  except for the $E=k_z$ 
topologically bound state   which is  anihilated by $Q$, come  in pairs,
$\psi$ and ${Q}\psi$, with opposite energy. If this pairing were really valid,  then  all  terms, again with the exception of   the unpaired bound state, would cancel in (\ref{EQ:oldedge}).  The formal $E\leftrightarrow-E$ symmetry suggested  by the existence of $Q$ is illusory, however. In order for $H_{\rm Dirac}$ to possess a well defined eigenvalue problem we  must impose some self-adjoint boundary conditions on the eigenfunctions. If $\psi$ obeys these boundary conditions, then, in  general, $Q\psi$ will not, and the actual density of states is  asymmetric across $E=0$. If,  as was the case in the days when \cite{anderson-morel,cross1} were written, one does not know of the topologically  bound edge state, or 
is   unaware of the fatuous nature of the symmetry implied by operators such as  $Q$, then it might   seem  that the only  contributions to the edge current  come from   the $O(\Delta/\epsilon_F)$  particle-hole asymmetry  that arises from the curvature of the  dispersion relation near the Fermi surface---exactly as  claimed in \cite{anderson-morel,cross1}.  

In conclusion, we see that when solved  exactly  the  BdG formalism   produces a mass flow and angular momentum that coincides  that obtained from  the Cooper-pair wave function: there is no  $\Delta/\epsilon_F$ suppression, and the  ground-state intrinsic angular momentum is $\frac12 \hbar$ per particle.  

\section{Acknowledgements}

This work was supported by the National Science Foundation under grant DMR-06-03528.  We would like to thank Rahul Roy and Shizhong Zhang for the original suggestion  that  we investigate $p_x+ip_y$ fermions in a harmonic trap, and for discussions in the early stages of the project..


\begin{thebibliography}{99}

\bibitem{kita1} T.~Kita, J.\ Phys.\ Soc.\ Japan {\bf 65}, 664 (1996).
\bibitem{leggett_book} A.~J.~Leggett, {\it Quantum Liquids: Bose Condensation and Cooper Pairing in Condensed-Matter Systems\/}, (Oxford University Press 2006) p376.
\bibitem{maeno_review} For a review, see: A.~P.~Mackenzie, Y.~Maeno, Rev.\ Mod.\ Phys.\ {\bf 75}, 657 (2003). 
\bibitem{jin} C.~A.~Regal, M.~Greiner,  D.~S.~Jin, Phys.\ Rev.\ Lett.\ {\bf  92}, 040403 (2004).

\bibitem{anderson-morel} P.~W.~Anderson, P.~Morel, Phys.\ Rev.\ {\bf 123}, 1911 (1961).
\bibitem{cross1} M.~C.~Cross, J.\ Low Temp.\ Phys.\  {\bf 21}, 525 (1975).
\bibitem{McClure} M.~G.~McClure, S.~Takagi, Phys.\ Rev.\ Lett.\ {\bf 43}, 596 (1979).
\bibitem{kita2} T.~Kita, J.\ Phys.\ Soc.\ Japan, {\bf 67}, 216 (1998). 
\bibitem{stone-nb} A {\it Mathematica}$^{\rm TM}$ notebook containing suitable code can be obtained by contacting  the authors.
\bibitem{volovik-edge} G.~E.Volovik, Pis'ma V Zh.\
Eksp.\ Teor.\ Fiz.\ {\bf 66}, 492-6 (1997) (JETP Letters, {\bf 66}, 522 (1997)).
\bibitem{read-green} N.~Read, D.~Green, Phys.\ Rev.\
B {\bf 61}, 10267 (2000).
\bibitem{mermin-muzikar} N.~D.~Mermin, P.~Muzikar, Phys.\ Rev.\  B {\bf 21}, 980 (1980).
\bibitem{ishikawa0} M.~Ishikawa, Prog.\ Theor.\ Phys.\ {\bf 57}, 1836  (1976); {\it Ibid\/}.\ {\bf 63}, 338 (1980).
\bibitem{ishikawa} M.~Ishikawa,  K.~Miyaki, T.~Usui,  Prog.\ Theor.\ Phys.\ {\bf 63}, 1083  (1980).
\bibitem{goldstone-wilczek} J.~Goldstone, F.~Wilczek
Phys.\ Rev.\ Lett.\ {\bf 47}, 986
(1981). 
\bibitem{stone-garg-muzikar} A.~Garg, P.~Muzikar,  M.~Stone,  Phys.\ Rev.\ Lett.\  {\bf 55}, 2328  (1985).
\bibitem{volovik-flow} G.~E.~Volovik, Pis'ma ZhETF,  {\bf 43}  428 (1986).
\bibitem{balatskii-volovik-konyshev} A. V. Balatskii , G. E. Volovik, V. A. Konyshev,
Zh.\ Eksp.\ Teor.\ Fiz.\ {\bf 90}, 2038  (1986) (Sov.\ Phys.\ JETP 63, No.6 (1986)).
\bibitem{stone-gaitan} F.~Gaitan, M.~Stone, Annals of Physics (NY) {\bf 178}, 89 (1987).
\bibitem{volovik-texture} G.~E.~Volovik, Pis'ma ZhETF , {\bf 61}  935 (1995).
\bibitem{stone-roy} M.~Stone, R.~Roy, Phys.\ Rev.\ B {\bf 69}, 184511 (2004).
\end{thebibliography}
\end{document}